\documentclass[twocolumn]{aastex631}
\usepackage{multirow}

\graphicspath{{./}{}}

\begin{document}
\font\myfont=cmr12 at 20pt
\title{{\myfont Solar Anti-Hale Bipolar Magnetic Regions:\\A Distinct Population with Systematic Properties}}

\correspondingauthor{Andrés Muñoz-Jaramillo}
\email{amunozj@boulder.swri.edu}

\author{Andrés Muñoz-Jaramillo}
\affiliation{Southwest Research Institute\\
1050 Walnut Street, \#300\\
Boulder, CO 80302, USA}

\author{Benjam\'in Navarrete}
\affiliation{Departamento de Astronom\'ia, Universidad de Chile, Casilla 36-D, Santiago, Chile}


\author{Luis E. Campusano}
\affiliation{Departamento de Astronom\'ia, Universidad de Chile, Casilla 36-D, Santiago, Chile}




\begin{abstract}
Besides their causal connection with long and short-term magnetic variability, solar bipolar magnetic regions are our chief source of insight into the location, size, and properties of large-scale toroidal magnetic structures in the solar interior.  The great majority of these regions ($\approx$95\%) follow a systematic east-west polarity orientation (Hale's law) that reverses in opposite hemispheres and across even and odd cycles.  These regions also present a systematic north-south polarity orientation (Joy's law) that helps build the poloidal field that seeds the new cycle.   Exceptions to Hale's law are rare and difficult to study due to their low numbers.   Here we present a statistical analysis of the inclination (tilt) with respect to the equator of Hale vs.\ Anti-hale regions spanning four solar cycles, considering two complementary tilt definitions adopted in previous studies.  Our results show that Anti-Hale regions belong to a separate population than Hale regions, suggesting a different originating mechanism.  However, we find that Anti-Hale region tilts present similar systematic tilt properties and similar latitudinal distributions as Hale regions, implying a strong connection between the two. We see this as evidence that they belong to a common toroidal flux system.  We speculate that Anti-hale regions originate from poloidal field sheared and strengthened "on-the spot" after the emergence of hale regions with very strong poloidal contribution.  Thus, they are not in contradiction with the idea of largely coherent toroidal flux systems inside the solar interior.
\end{abstract}



\section{Introduction} \label{sec:intro}

Bipolar magnetic regions (BMRs) are regions of strong magnetic fields that emerge to the solar photosphere the solar photosphere, acting as the main source of space weather events.  They typically form sunspots, where strong magnetic field bundles cross the photosphere \citep{vanDriel2015}, appearing as dark contrast regions on the solar photosphere in visible light.  The number of BMRs present at any given moment on the solar surface waxes and wanes during the solar cycle.  Although we cannot observe the solar interior where BMRs originate, their observation give us insight into the dynamical behavior of the large-scale internal solar magnetic field.

\begin{figure*}[ht!]
\includegraphics[width=18cm,scale=0.5]{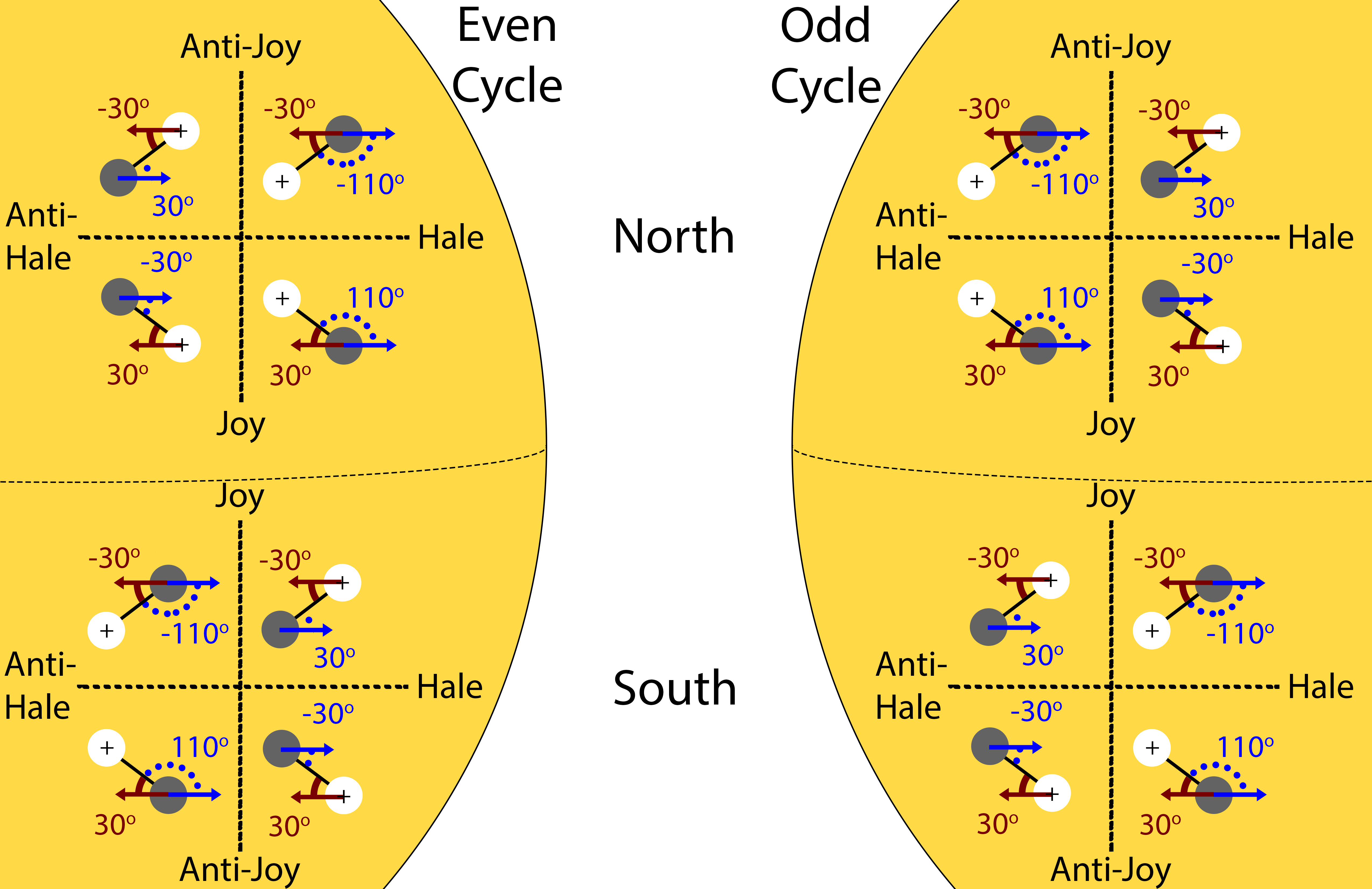}
\caption{Schematic description of the different combinations of tilt and hale orientations and how they translate to angles depending on the reference frame used.  Blue angles use the negative polarity as the reference frame and go between $-180^o$ and $180^o$.  This definition is better suited to highlight the difference between Hale (HH) and Anti-Hale (AH) regions.    Red angles use the leading polarity as the reference frame and go between $-90^o$ and $90^o$.  This definition is better suited to highlight the difference between Joy and anti-Joy regions. \label{fig:tilt_types}}
\end{figure*}

\begin{figure*}[ht!]
\includegraphics[width=\textwidth]{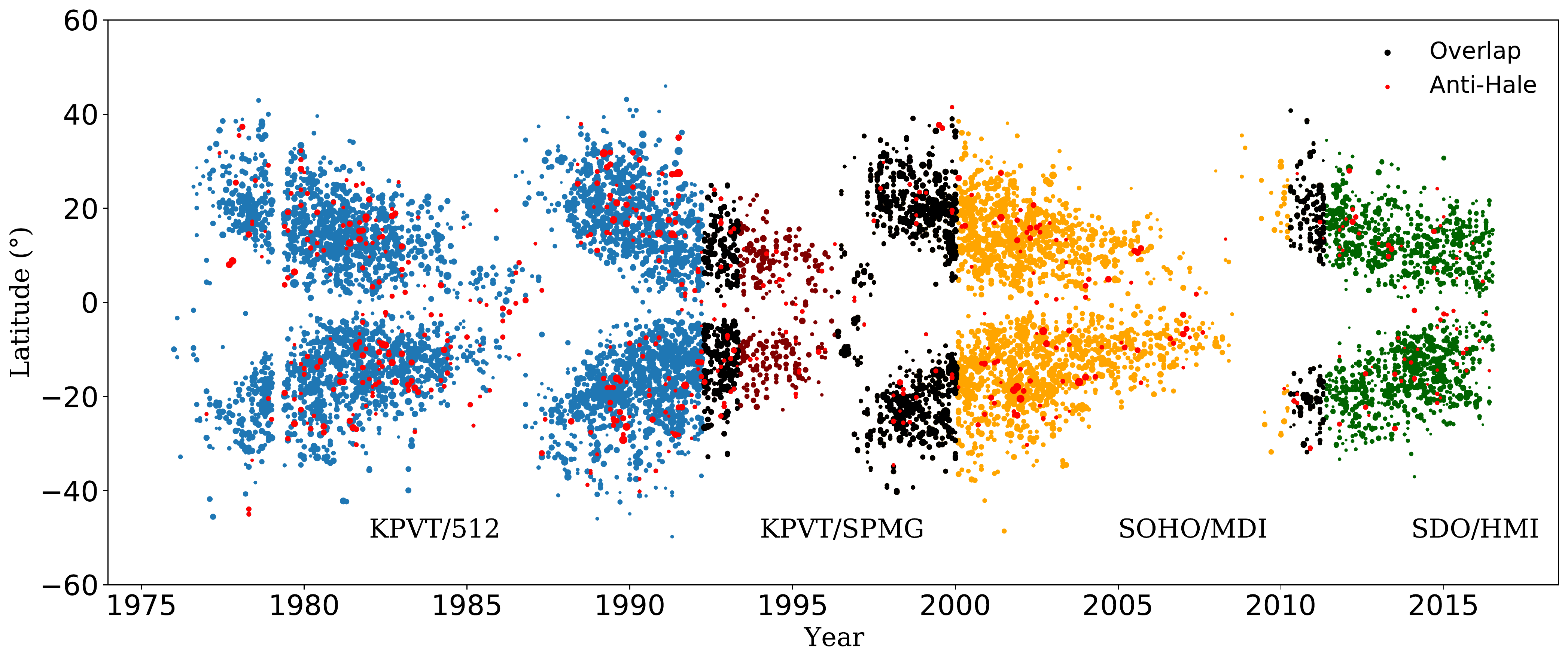}
\caption{Solar Latitude of active regions as a function of time (Butterfly diagram) for the last four solar cycles. Blue data is from KPVT/512, purple for KPVT/SPMG, orange for SOHO/MDI and green color for SDO/HMI measurements. The overlap regions between instruments are plotted in black color.  Anti-Hale regions are shown in red.  \label{fig:butterfly}}
\end{figure*}

Three main empirical laws govern populations of active regions: Spörer's Law \citep{carrington1858} states that the latitude of emerging sunspots decreases with time as the solar cycle goes forward (from mid-latitudes towards the equator),  giving rise to the time vs.\ latitude plot called the butterfly diagram (see Fig.~\ref{fig:butterfly}). Hale's law \citep{hale1925}, which establishes that BMRs present a systematic East-West magnetic polarity orientation for a given solar hemisphere -- for example, if the positive pole of the BMR trails behind the negative pole at the northern solar hemisphere in the east-west direction, then the negative pole will be behind the positive in the south in the same direction.  The leading polarity of each hemisphere is reversed for odd and even cycles.  Finally, Joy's law \citep{hale-etal1919} expresses that BMRs tend to have a systematic North-South orientation, with the leading (trailing) polarity being closer to the equator (pole).  This law is typically quantified through the measurement of a BMR's tilt (i.e.\ the angle between a line that passes between the center off a BMR's polarities and a line parallel to the solar equator).   Regions that follow this law are typically refered to as Hale (HH) regions.


There is a small fraction of BMRs that violate Hale's law and its hemispheric rule \citep{richardson1948}: these ``anti-Hale (AH)'' regions have an opposite East-West polarity orientation than most of the BMRs in the respective hemisphere.   These regions comprise between 5-10\% of all BMRs \citep{richardson1948, wang-sheeley1989, khlystova-sokoloff2009, mcclintock-etal2014, stenflo-kosovichev2012, li2018}.  In spite of their relatively small numbers, AH regions provide important clues to the structure of the internal magnetic field.


Previously, \cite{stenflo-kosovichev2012} argued that AH regions are a distinct population, rather than anomalies of the HH groups.  They suggest that the appearance of AH regions in similar latitudes that HH regions rules out the possibility of well-defined, coherent toroidal flux systems as a source of all active regions. On the other hand, \cite{mcclintock-etal2014} hypothesized that AH regions are part of a continuous distribution of regions that stems from the HH population, a picture that more consistent with the existence of a largely coherent toroidal flux system.  Here we provide further evidence that AH regions comprise a different population than HH regions, implying that their origin is different than that of HH. However, we still find a strong connection between some of their statistical properties, which we take as evidence of a causal connection between them.



In Section \ref{sec:tilt definitions} we introduce the tilt definitions used in this work. Sections \ref{sec:data} and  \ref{sec:flux calibrations} describe the data and the determination of flux cut-offs and calibration factor respectively. In Section \ref{sec:hale fractions} we determine the Hale and anti-Hale fractions by hemisphere. Section \ref{sec:general combination} describes the combination of BMRs from different hemispheres and Cycles.
In section \ref{sec:joyslaw} we analyze to what extent do Hale and Anti-Hale BMRs follow Joy's Law.  In section \ref{sec:globalDist} we report the optimal analytic fit to the distribution of Hale and anti-Hale region tilts. In section \ref{sec:HaleDev} we demonstrate that Hale and anti-Hale region tilts cannot be described using a unique distribution. Finally, in Section \ref{sec:origin of AH regions} we discuss the possible origin of the Anti-Hale regions and in Section \ref{sec:conclusions} we provide a summary with the main conclusions.

\section{Absolute vs. Relative BMR Magnetic tilt definitions}
\label{sec:tilt definitions}

There are two main ways of defining the tilt of a BMR that showcase different aspects of BMR inclination (see Fig.~\ref{fig:tilt_types}).  They differ on the way BMR poles are chosen as references.  They have different flavors in the way they represent the combination of HH \textbf{vs.}AH and Joy (leading polarity closer to the equator) vs.\ anti-Joy (leading polarity closer to the pole).

One possibility is to always use the same polarity (typically the negative polarity) as the location of the reference frame and the other polarity (typically the positive) to determine the tilt angle \citep[see for example][]{li-ulrich2012, mcclintock-etal2014}.   Under this definition, tilt angles vary between $-180^o$ and $180^o$.   This reference frame (and associated angles) are shown using blue ink in Fig.~\ref{fig:tilt_types}.  Under this definition, HH vs.\ AH regions will separate into angles with a magnitude larger vs.~smaller than $90^o$ (depending on the hemisphere and cycle).   Joy vs. anti-Joy are in turn differentiated by a change of angle sign for a given hemisphere and cycle.  This approach, which we refer to from now on as the \textit{\textbf{``absolute tilt"}}, enables the simultaneous analysis of Hale orientation and Joy tilt angle, while treating all BMRs as part of a continuous distribution of angles.  It contains the implicit assumption that HH and AH regions belong to a continuously distributed population.

The second possibility is to always use the leading polarity (regardless of its polarity) as the location of the origin and the other polarity (the trailing polarity) to determine the tilt angle \citep[see for example][]{wang-sheeley1989, howard1991, stenflo-kosovichev2012}.   Under this definition, tilt angles vary between $-90^o$ and $90^o$.   This reference frame (and associated angles) are shown using red ink in Fig.~\ref{fig:tilt_types}.  Under this definition, HH vs. AH regions have to be explicitly separated since there is no information in the tilt angle that conveys Hale orientation.   Joy vs. anti-Joy BMRs are very easily differentiated because Joy (anti-Joy) regions always have positive (negative) angles in the Northern hemisphere (regardless of the cycle) -- signs reverse for the Southern hemisphere.  This approach, which we refer to from now on as the \textit{\textbf{``relative tilt"}}, focuses exclusively on the relative position of BMR leading and trailing polarities.  It contains the implicit assumption that HH and AH regions belong to different populations.

To this date, analyses of tilt and Hale-orientation have chosen one approach or the other. We speculate that it is not accidental that studies that use the absolute tilt definition conclude that BMRs arise from a continuous distribution of angles, whereas studies that use the relative tilt definition conclude that HH and AH populations are different.   The aim of this study is to determine the significance limit at which we can reject the null hypothesis that the HH and AH regions constitute a single population and this is only possible to do if both definitions of tilt are studied simultaneously.

As done by \cite{howard1991}, we compute the magnetic tilt angles as:

\begin{equation}
    \tan(\gamma) = \Delta\lambda/(\Delta\phi \cos(\bar{\lambda})) \label{eq:tilt_def}
\end{equation}

Where $\bar{\lambda}$ is the midpoint latitude of the polarity pair, $\Delta\lambda$ and $\Delta\phi$ are the differences between latitude and longitude respectively.  For the absolute tilt calculation we set the negative polarity as the reference polarity and use the arc-tangent function with two arguments so that the resulting angles go between $-180^o$ and $180^o$.   For the relative tilt calculation we use the leading polarity as reference and use the single argument arc-tangent so that the resulting angles go between $-90^o$ and $90^o$.  We ensure that tilt angles are positive (negative) for Joy regions in the Northern (Southern) hemispheres.   \textit{For both relative and absolute tilts we use the division shown in Figure \ref{fig:tilt_types} to separate our data into Hale and AH regions depending on cycle and hemisphere}.


\section{Data}
\label{sec:data}

We use data from two ground-based and two spaced-based instruments.   Our ground-based instruments are part of the Kitt Peak Vacuum Telescope survey.  The first instrument is the 512 Diode Array Magnetograph \citep[from 1976 to 1993][]{livingston-etal1976} and the second the solar Spectromagnetograph \citep[SPMG, from 1992 to 2001][]{jones1992}. Both instruments obtained images at a cadence of one image per day. The 512 instrument has a resolution of 1" per pixel; SPMG has a resolution of 1.15" per pixel. Our space-based instruments are the Michelson Doppler Imager \citep[MDI;][1996-2010]{scherrer-etal1995}, on board the Solar and Heliospheric Observatory (SOHO) collected 136,839 magnetograms from 1996 to 2011 at a 96 minutes cadence and a resolution of 2" per pixel.  We also use the data from the Helioseismic and Magnetic Imager \citep[HMI;][2010-present]{scherrer-etal2012}, on board the Solar Dynamics Observatory (SDO), which produces magnetograms with a resolution of 0.5" per pixel at a 45 second cadence.

\begin{figure}[ht!]
\includegraphics[scale=0.445]{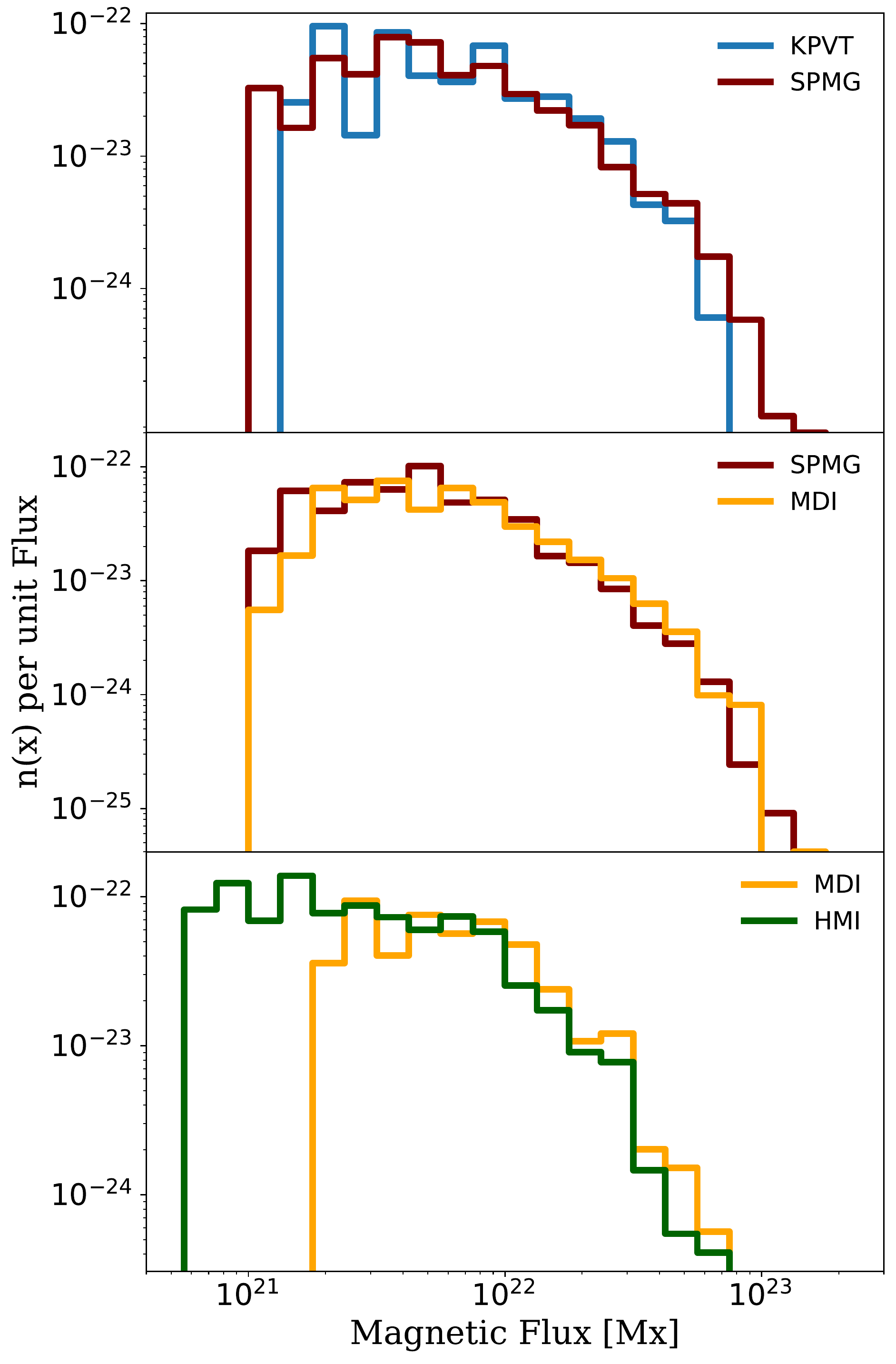}
\caption{Empirical BMR flux distributions for the overlapping periods between instruments.  Between KPVT/512 and KPVT/SPMG from April 1992 to April 1993, SPMG-MDI from May 1996 to December 1999, and MDI-HMI from April 2010 to April 2011. Blue line represents the distribution of KPVT/512, KPVT/SPMG in orange, SOHO/MDI in green, and SDO/HMI in red. \label{fig:overlap_flux}}
\end{figure}

Our BMR dataset consists of 9,243 individually tracked BMRs, obtained using the Bipolar Active Region Detection (BARD) code described by \citet{paper_databases}.   Here we use a single tilt measurement per BMR, at the moment of maximum flux, to avoid folding the time evolution of a BMR tilt during a BMR's lifetime into our analysis.  For each BMR we measure the flux, area, and flux-weighted latitude and longitude centroids of the positive and negative polarities.  Data is shown in Figure \ref{fig:butterfly}.  Our data covers four solar cycles (21-24) as seen by the four different instruments (KPVT/512, KPVT/SPMG, SOHO/MDI, and SDO/HMI).


\subsection{Combination of data from different instruments}


\begin{figure*}[ht!]
\includegraphics[width=18cm,scale=0.5]{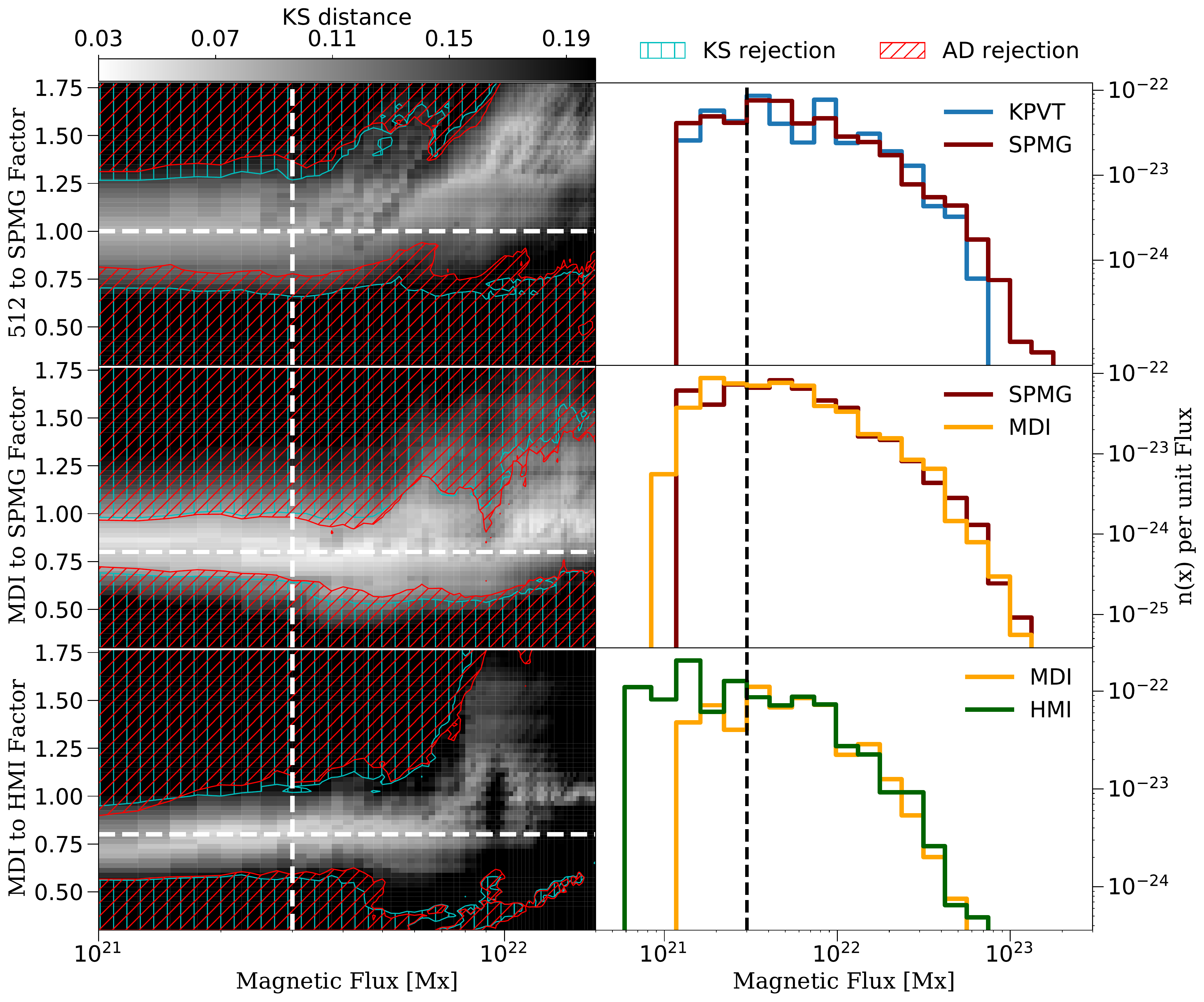}
\caption{Two dimensional optimization of calibration factor and flux cut for the overlapping periods between instruments.  \textit{Left:}  Brightness indicates the KS distance between distributions.   Blue vertical hatches (diagonal red hatches) indicate the region where the Kolmogorov-Smirnov (Anderson-Darling) statistics rejects the hypothesis of the two populations belonging to the same distribution.  White dashed lines show the optimal calibration factor (horizontal) and universal magnetic flux cutoff (vertical) that maximize the similarity between distributions.  The only distribution that needs to be calibrated is SOHO/HMI, with a factor of 0.8, which makes it homogeneous both with KPVT/SPMG and SDO/HMI.  We find that a universal flux cutoff of $3\times 10^{21}$Mx is an optimal compromise between maximizing distribution homogeneity and minimizing the number of BMRs that have to be rejected.   Right:  Homogenized distributions showing the universal cutoff of $3\times 10^{21}$Mx with a black dashed line.  MDI flux has been applied a 0.8 correction factor. \label{fig:optimize}}
\end{figure*}

Figure \ref{fig:overlap_flux} shows the distributions of net BMR magnetic flux during overlap intervals (512-SPMG, SPMG-MDI and MDI-HMI). In each of these histograms we observe that there is general agreement for strong BMRs, with the discrepancies arising for small objects.   This discrepancy arises from an artificial cut-off introduced by the difference in sensitivity and resolution across magnetographs (very clear in the case of the MDI vs.\ HMI comparison).  Ensuring that the populations across instruments sample a similar range of fluxes is important because tilt scatter is dependent on flux \textbf{\citep{stenflo-kosovichev2012,jiang-etal2014}}. Furthermore, due to the fact that small objects are more numerous, they have a sizable effect on tilt statistics.

Given that our tilt angles are calculated using the flux weighted centroid of the positive and negative regions in a BMR, we believe our tilt measurements are largely independent of instrument cross-calibration issues and are internally consistent.  Because of this a rough calibration factor between instruments, coupled with a universal flux cut-off across instruments, are sufficient to ensure that we are comparing populations with similar flux statistics.  We demonstrate quantitatively that this is the case in Section \ref{sec:general combination}.

\section{Determination of flux cut-offs and approximate calibration factors}
\label{sec:flux calibrations}
We take advantage of the time-overlap between instruments to derive an approximate flux calibration, as each of the instruments has its own resolution and observational systematic issues.   To find the optimal calibration factor, we minimize the two-sample Kolmogorov-Smirnov (K-S) distance between the flux distribution of the overlapping period for each pair of instruments.   We do this after applying a calibration factor to one of the instruments and a common flux cutoff for both the target and calibrated instrument.   This form of calibration was used by \cite{munoz-etal2015a} to calibrate sunspot group areas.  Calibrating the strong end of the distribution (i.e.\ the big BMRs that all instruments see well) is the best way of homogenizing our data because small BMRs are systematically under-observed due to cadence and sensitivity issues.  Furthermore, these systematic issues are different for each instrument.  Introducing a cut-off allows the fit to focus on the large BMRs that all instruments can see equally well.


Figure \ref{fig:optimize} shows the result of the 2-dimensional optimization (calibration factor plus a common cutoff after calibration). Dark (light) regions indicate a large (small) K-S distance. Hatched areas indicate parameters for which the K-S test and the Anderson-Darling (A-D) tests reject the hypothesis that BMRs observed by different instruments belong to the same population.  There are different combinations of factors and flux cuts that optimize the match between distributions. However, there is a trade off between K-S distance and cutoff: the more regions we leave out, the easier it is to find a fit.   However, leaving out too many regions is counter productive because it will reduce the number of regions we can analyse. We find two main optimal regions:  a region around a cutoff of $3\times 10^{21}$ and another around $10^{22}$.   We pick the cutoff value of $3\times 10^{21}$ in order to retain as many BMRs as possible for our analysis.   This translates into a calibration factor around 1 for KPVT/512 vs KPVT/SPMG, 0.8 for SOHO/MDI vs for KPVT/SPMG, and 0.8 for SOHO/MDI vs SOHO/HMI. Thus a 0.8 factor applied to MDI is sufficient to get approximate consistency in the flux distribution for all instruments and all cycles.   This factor in good agreement ($1/0.8 = 1.25$) with the 1.3 factor found by \cite{liu2012} to calibrate HMI to MDI for strong magnetic fields, considering that the focus of our analysis are BMRs containing the strongest magnetic fields in the solar photosphere.

\begin{figure*}[ht!]
\includegraphics[width=\textwidth]{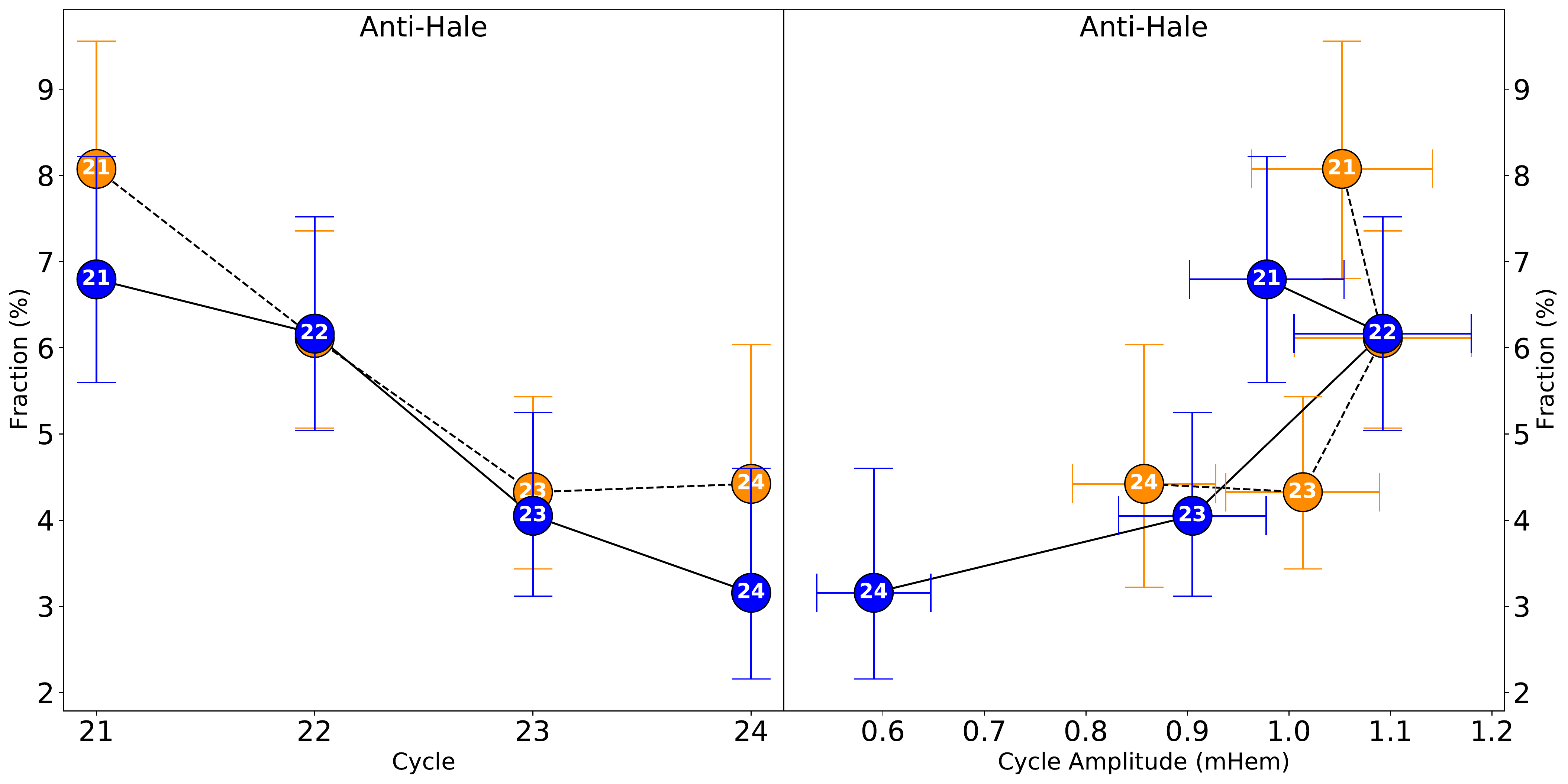}
\caption{Anti-Hale fraction of all BMRs. As a function of solar cycle number \textit{(Left)} and solar cycle amplitude \textit{(Right)}.  The Northern (Southern) hemisphere is shown using blue (orange) markers. \label{fig:porcentajes}}
\end{figure*}



Once the calibrated factor is applied, we can see a better agreement between the flux distributions of the different instruments for strong BMRs (see right panels of Fig.~\ref{fig:optimize} vs. panels in Fig.~\ref{fig:overlap_flux}).  We use $3\times 10^{21}$Mx as a universal cutoff for all instruments (after applying the 0.8 factor to MDI), shown in the right column of Fig.~\ref{fig:optimize} as a vertical black dashed line.  Any BMR with less flux than this cutoff is removed from our analysis.   We discard $1,232$ BMRs leaving $\sim 87\%$ from the initial $9,243$ BMRs in the database. Additionally, we only include BMRs from a single instrument in each overlap.  This ensures that we do not count each BMR more than once, leaving $7,511$ flux-normalized BMRs for use in our analysis of tilts.


\begin{table*}
\centering
\caption{Numbers and fractions of Hale and anti-Hale BMRs per cycle and hemisphere.}
\begin{tabular*}{0.8\linewidth}{@{\extracolsep{\fill}}lcrrcrc}
\hline
\hline
\multirow{2}{*}{Cycle} & \multirow{2}{*}{Hem.} & \multicolumn{1}{c}{\multirow{2}{*}{\#}} & \multicolumn{2}{c}{Hale} & \multicolumn{2}{c}{anti-Hale} \\ \cline{4-5} \cline{6-7}
 & & \multicolumn{1}{c}{} & \multicolumn{1}{c}{\#} & \% &  \multicolumn{1}{c}{\#} & \% \\ \hline
\multirow{2}{*}{21} & N & 1001 & 933 & 93 &  68 & 7 \\
 & S & 1065 & 979 & 92 &  86 & 8 \\ \hline
\multirow{2}{*}{22} & N & 1022 & 959 & 94 & 63 & 6 \\
 & S & 1194 & 1121 & 94 &  73 & 6 \\ \hline
\multirow{2}{*}{23} & N & 938 & 900 & 96 & 38 & 4 \\
 & S & 1133 & 1084 & 96 &  49 & 4\\ \hline
\multirow{2}{*}{24} & N & 570 & 552 & 97 &  18 & 3 \\
 & S & 588 & 562 & 96 & 26 & 4 \\ \hline
Total: & \multicolumn{1}{l}{} & 7511 & 7090 & 94 & 421 & 6
\end{tabular*}
\label{tab:hale fractions}
\end{table*}

\section{Hale and anti-Hale fractions by cycle and hemisphere}
\label{sec:hale fractions}

Table \ref{tab:hale fractions} and Figure \ref{fig:porcentajes} show the numbers and fractions of the Hale and AH sunspots for each hemisphere and cycle. Overall, the fraction of AH sunspots in our data is $5.61\%$ which is consistent with previous estimates \citep{richardson1948, wang-sheeley1989, khlystova-sokoloff2009, stenflo-kosovichev2012}.  Other studies \citep{mcclintock-etal2014, li2018} have reported a larger fraction of AH BMRs ($~8\%$).  This discrepancy is likely driven by the inclusion of small objects in other surveys, which invariably dominate BMR statistics due the fact that smaller regions are more numerous.  \cite{li2018} reports an AH fraction of $~8\%$, but applying a flux cutoff of $3\times 10^{21}$ to their Figure 10, like the one we use in this work, reduces the fraction to $~6\%$.  This does not invalidate in any way their results.  It simply highlights the importance of understanding the consequences of working with objects near a survey's threshold of detection.  It also highlights the difficulty of comparing surveys with different (and unspecified) detection thresholds.

There seems to be a downward trend in the proportion of AH BMRs from cycle to cycle in our observed cycles (21-24; see left panel of Fig.~\ref{fig:porcentajes}).  This trend is also visible in the results of \cite{mcclintock-etal2014} and \cite{li2018}.  It is possible that there is a relationship between cycle amplitude and how many AH regions it produces (see the right panel of Fig.~\ref{fig:porcentajes}): the stronger the cycle, the higher the proportion of AH regions.  This relationship does not seem to be very strong, but it is still significant after estimating the Wilson uncertainty interval for a binomial distribution \citep{Brown2001}. As will be explained in Section \ref{sec:origin of AH regions}, we speculate that this hints to connection between AH regions and preceding HH regions with a strong poloidal field component.


\begin{table*}[ht!]
\caption{Kolmogorov-Smirnov test of the progressive combination of cycle and hemispheric BMRs into a unified set.}
\label{tab:ks_before_merge}
\centering
\begin{tabular}{lcccccccccc}
\hline
\hline \\
\multicolumn{11}{c}{\textbf{\Large Absolute Tilt}} \\
\multirow{3}{*}{\textbf{Population}} & \multirow{3}{*}{\textbf{Hemisphere}} & \multicolumn{1}{l}{\multirow{3}{*}{\textbf{Cycle}}} & \multicolumn{1}{l}{\multirow{3}{*}{\textbf{\textit{D}-statistic}}} & \multicolumn{1}{l}{\multirow{3}{*}{\textbf{\textit{p}-value}}} &  &
\multicolumn{1}{l}{\multirow{3}{*}{\textbf{\textit{D}-statistic}}} & \multicolumn{1}{l}{\multirow{3}{*}{\textbf{\textit{p}-value}}} & &
\multicolumn{1}{l}{\multirow{3}{*}{\textbf{\textit{D}-statistic}}} & \multicolumn{1}{l}{\multirow{3}{*}{\textbf{\textit{p}-value}}} \\
 & \multicolumn{1}{l}{} & \multicolumn{1}{l}{} & \multicolumn{1}{l}{} & \multicolumn{1}{l}{} \\
 & \multicolumn{1}{l}{} & \multicolumn{1}{l}{} & \multicolumn{1}{l}{} & \multicolumn{1}{l}{} \\ \hline
\multirow{4}{*}{Hale} & \multirow{2}{*}{N} & 21+23 & 0.024  & 0.94 & \multirow{2}{*}{$>$} & \multirow{2}{*}{0.023} & \multirow{2}{*}{0.78} & \multirow{4}{*}{$>$} & \multirow{4}{*}{0.018} & \multirow{4}{*}{0.54}\\ \cline{3-5}
&  & 22+24 & 0.089 & 0.01 \\ \cline{2-8}
& \multirow{2}{*}{S} & 21+23 & 0.049 & 0.16 & \multirow{2}{*}{$>$} & \multirow{2}{*}{0.039} & \multirow{2}{*}{0.11}\\ \cline{3-5}
&  & 22+24 & 0.049 & 0.16 \\ \hline
\multirow{4}{*}{Anti-Hale} & \multirow{2}{*}{N} & 21+23 & 0.100 & 0.94 & \multirow{2}{*}{$>$} & \multirow{2}{*}{0.107} & \multirow{2}{*}{0.61} & \multirow{4}{*}{$>$} & \multirow{4}{*}{0.077} & \multirow{4}{*}{0.53}\\ \cline{3-5}
&  & 22+24 & 0.100 & 0.94 \\ \cline{2-8}
& \multirow{2}{*}{S} & 21+23 & 0.112 & 0.77 & \multirow{2}{*}{$>$} & \multirow{2}{*}{0.077} & \multirow{2}{*}{0.745}\\ \cline{3-5}
&  & 22+24 & 0.114 & 0.41 \\ \\
\hline
\hline\\
\multicolumn{11}{c}{\textbf{\Large Relative Tilt}} \\
\multirow{3}{*}{\textbf{Population}} & \multirow{3}{*}{\textbf{Hemisphere}} & \multicolumn{1}{l}{\multirow{3}{*}{\textbf{Cycle}}} & \multicolumn{1}{l}{\multirow{3}{*}{\textbf{\textit{D}-statistic}}} & \multicolumn{1}{l}{\multirow{3}{*}{\textbf{\textit{p}-value}}} &  &
\multicolumn{1}{l}{\multirow{3}{*}{\textbf{\textit{D}-statistic}}} & \multicolumn{1}{l}{\multirow{3}{*}{\textbf{\textit{p}-value}}} & &
\multicolumn{1}{l}{\multirow{3}{*}{\textbf{\textit{D}-statistic}}} & \multicolumn{1}{l}{\multirow{3}{*}{\textbf{\textit{p}-value}}} \\
 & \multicolumn{1}{l}{} & \multicolumn{1}{l}{} & \multicolumn{1}{l}{} & \multicolumn{1}{l}{} \\
 & \multicolumn{1}{l}{} & \multicolumn{1}{l}{} & \multicolumn{1}{l}{} & \multicolumn{1}{l}{} \\ \hline
\multirow{4}{*}{Hale} & \multirow{2}{*}{N} & 21+23 & 0.024 & 0.94 & \multirow{2}{*}{$>$} & \multirow{2}{*}{0.022} & \multirow{2}{*}{0.78} & \multirow{4}{*}{$>$} & \multirow{4}{*}{0.020} & \multirow{4}{*}{0.54}\\ \cline{3-5}
&  & 22+24 & 0.07 & 0.04 \\ \cline{2-8}
& \multirow{2}{*}{S} & 21+23 & 0.041 & 0.35 & \multirow{2}{*}{$>$} & \multirow{2}{*}{0.039} & \multirow{2}{*}{0.11}\\ \cline{3-5}
&  & 22+24 & 0.054 & 0.22 \\ \hline
\multirow{4}{*}{Anti-Hale} & \multirow{2}{*}{N} & 21+23 & 0.166 & 0.46 & \multirow{2}{*}{$>$} & \multirow{2}{*}{0.108} & \multirow{2}{*}{0.61} & \multirow{4}{*}{$>$} & \multirow{4}{*}{0.077} & \multirow{4}{*}{0.53}\\ \cline{3-5}
&  & 22+24 & 0.230 & 0.40 \\ \cline{2-8}
& \multirow{2}{*}{S} & 21+23 & 0.112 & 0.77 & \multirow{2}{*}{$>$} & \multirow{2}{*}{0.114} & \multirow{2}{*}{0.41}\\ \cline{3-5}
&  & 22+24 & 0.311 & 0.04 \\
\end{tabular}
\end{table*}

\section{Combination of BMR's from different hemispheres and Cycles}
\label{sec:general combination}
Polarity orientation is opposite for the Northern and Southern hemispheres and reverses with every new cycle.  This means that the absolute (relative) tilt angles that represent the different combinations of Joy vs.\ anti-Joy and Hale vs.\ AH orientations can adopt different sets of values for different hemispheres and odd/even cycles; 4 in the case of the absolute tilt and 2 in the case of the relative tilt (see Figure \ref{fig:tilt_types}). In order to take full advantage of four cycle's worth of BMRs, we first verify that BMR tilts in different cycles can be considered to belong to the same population and then proceed to fold them into a single population by modifying their tilt angles to become what they would be if hemispheric and cyclic reversals would not take place.

Table \ref{tab:ks_before_merge} shows the results of the Kolmogorov-Smirnov (KS) test for a hierarchical merging of our BMRs into a single population.  We perform this merging separately for Hale and anti-Hale BMRs.   Each pair of columns named \textit{D-Statistic} and \textit{p-value} indicate the maximum difference between the empirical cumulative distribution functions of both populations \textit{D-Statistic} and the significance with we can reject the null hypothesis that they belong to the same population (1 - \textit{p-value})*100\%.

Merging BMR tilts belonging to different cycles and hemispheres requires the following manipulations (see Fig.~\ref{fig:tilt_types}):
\begin{itemize}
    \item In the case of relative tilt, all BMRs from the same hemisphere can be combined without manipulation and BMRs of different hemispheres can be combined after a sign change.
    \item In the case of the absolute tilt, tilt angles belonging to opposite hemispheres and subsequent cycles are consistent (i.e.\ see bottom-left and top-right of Fig.~\ref{fig:tilt_types}).  The other combination of cycle parity and hemisphere needs to be adjusted by 180$^o$ in order to make all angles consistent.
\end{itemize}

We first start by combining BMRs from the same hemisphere for odd (21+23) or even (22+24) cycles.  Overall, the null hypothesis that BMRs from the same hemisphere and odd/even cycles belong to the same population holds in almost all cases.  The only exception are 22N+24N for Hale regions and 22S+24S for AH regions.  We suspect that in these cases the null hypothesis is rejected due to the fact that our observations for cycle 24 are incomplete, so we may not have sufficient statistics to fully sample the tilt population (i.e.\ randomly drawing BMRs from the common distribution will, in some cases, result in different populations).  However, as we progressively merge all regions of the same hemisphere for all cycles (6th and 7th columns of Table \ref{tab:ks_before_merge}) and all Hale or AH regions (last two columns of Table \ref{tab:ks_before_merge}) the null hypothesis that BMRs from all cycles and hemispheres belong to the same population holds with a very high degree of significance.

\begin{figure*}[ht!]
\includegraphics[width=\textwidth, scale=0.7]{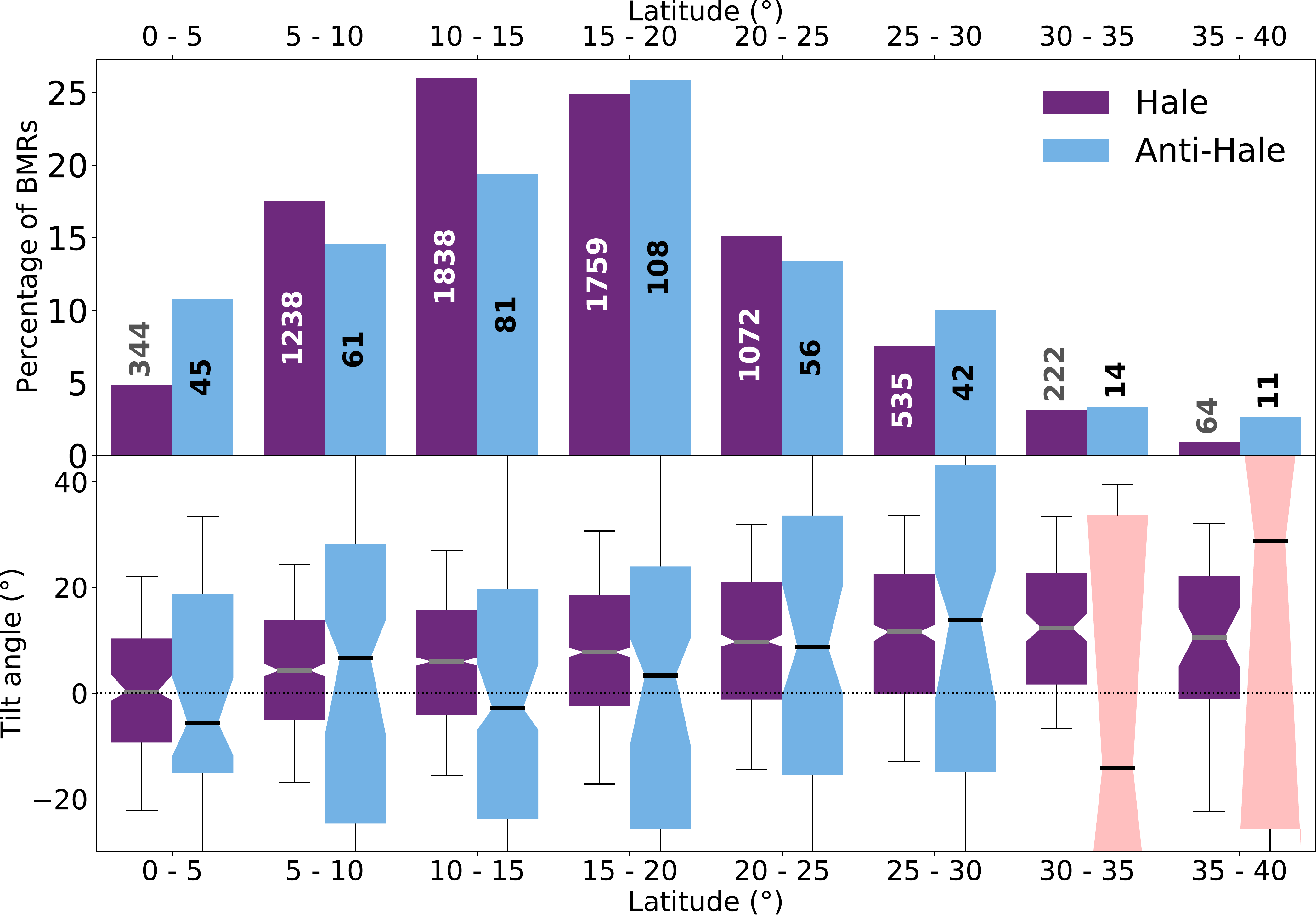}
\caption{\textit{Top:} Relative abundance of Hale (dark purple) and Anti-Hale (light blue) regions as a function of latitude.   The numbers on each column indicate the number of regions in a given bin.  \textit{Bottom:} Boxplot illustrating the distribution of tilt angles for each latitude.  Bold horizontal lines mark the median of the distribution.  The sloping notches indicate the uncertainty on the median estimation calculated using 1000 random subsets of the population.  The box indicates the extent of the 2nd and 3rd quartiles in the distribution and the whiskers mark the 5\% and 95\% percentiles.   Light-red denotes bins with so few points that no meaningful estimation of the median can be made. \label{fig:joyslaw}}
\end{figure*}

\section{Joy's Law for Hale and Anti-Hale BMRs} \label{sec:joyslaw}

The combination of all cycles and hemispheres gives us enough statistics to look at the dependence of tilt angle on latitude for AH BMRs in comparison to HH BMRs.   Figure \ref{fig:joyslaw} shows the relative abundance of HH and AH regions for different latitude bins, as well as box plots of HH and AH relative tilt angle distributions as a function of latitude.  Overall, the relative abundance of HH and AH regions as a function of latitude (top panel of Fig.~\ref{fig:joyslaw}) is very similar.  Performing a K-S test between the latitudes of HH and AH regions we find that we cannot reject the null-hypothesis that they are the same population with full confidence (p-value of 0.07).   We interpret this as evidence that the majority of AH regions are likely connected in some way with the flux that gives rise to HH regions.  The two distributions peak slightly differently, with the AH distribution peaking at a higher latitude than the HH distribution.  Additionally, there is a significant difference between the two distributions near the equator (0-5 bin).  We interpret this as evidence that some of the AH regions near the equator are likely normal HH regions belonging to the opposite hemisphere.

In terms of the latitudinal dependence of the distribution of tilts (bottom panel of Fig.~\ref{fig:joyslaw}).  The first clear difference between HH and AH regions is the spread in the distribution of tilt angles: for Hale regions 50\% of all tilt angles within each latitude bin are within 10$^o$ of the median, whereas for AH regions they are within 20$^o$.  HH tilts systematically increase with latitude and, thanks to the very good statistics, the median tilt can be determined with good certainty (the uncertainty interval indicated by the notches in the boxes is $\sim\pm 3^o$).  AH regions, on the other hand, present significantly more uncertainty in the position of their median ($\sim\pm 10^o$) due to their larger scatter and poorer statistics.  In the case of the 30-55$^o$ and 35-40$^o$ bins (denoted with a ligh-red color) the uncertainty is so large, that any measurement of the median is meaningless.  The uncertainty in AH medians makes it hard for us to place strong constraints on the dependence of AH tilt angle on latitude.  Nevertheless, it is very clear that there is a systematic hemispheric tendency of the same sign as the one presented by HH regions:  i.e.\ the leading polarity of AH regions also tends to be closer to the equator than the trailing polarity.  It also seems that the median tilt does increase with latitude starting int the 10-15$^o$ bin.  It is possible that the two first bins (0-5$^o$ \& 5-10$^o$) reflect the impact of regions belonging to the opposite hemisphere that emerge across the 0$^o$ line.


\begin{table*}[ht!]
\caption{Akaike weights and KS test of 7 analytic distributions for HH and AH regions.  Distributions are ordered according to the relative AIC for HH regions.  Out of these distributions, the best fit to HH (AH) regions is the T-student (von Mises) distributions (denoted using bold letters).}
\centering
\begin{tabular}{lcrrrrcrrrr}
\hline
\hline\\
\multicolumn{11}{c}{\textbf{\Large Model Selection and Goodness of Fit}}\\
\textbf{Fitted} & & \multicolumn{4}{c}{\textbf{Hale - Best: Student's T}} &  & \multicolumn{4}{c}{\textbf{Anti-Hale - Best: von Mises}}\\ \cline{3-6} \cline{8-11}
\textbf{Distribution} & & \textbf{$\Delta$AIC} & \textbf{AIC$_w$} & \textbf{K-S distance} & \textbf{K-S p} & & \textbf{$\Delta$AIC} & \textbf{AIC$_w$} & \textbf{K-S distance} & \textbf{K-S p}\\
\hline
Student's T \textbf{(HH)} & & \textbf{0} & \textbf{$\sim$1.00} & \textbf{0.015} & \textbf{0.07} & & 51 & $\sim$0.00 & 0.033 & 0.73 \\
Laplace & & 41 & $\sim$0.00 & 0.019 & 0.01 & & 79 & $\sim$0.00 & 0.042 & 0.44 \\
Logistic & & 181 & $\sim$0.00 & 0.026 & $\sim$0.00 & & 59 & $\sim$0.00 & 0.023 & 0.98 \\
von Mises \textbf{(AH)} & & 255 & $\sim$0.00 & 0.043 & $\sim$0.00 & & \textbf{0} & \textbf{$\sim$1.00} & \textbf{0.047} & \textbf{0.29} \\
Skew-Normal & & 687 & $\sim$0.00 & 0.059 & $\sim$0.00 & & 51 & $\sim$0.00 & 0.034 & 0.71 \\
Normal & & 909 & $\sim$0.00 & 0.071 & $\sim$0.00 & & 49 & $\sim$0.00 & 0.033 & 0.73 \\
Cauchy & & 1265 & $\sim$0.00 & 0.057 & $\sim$0.00 & & 190 & $\sim$0.00 & 0.084 & 0.01 \\
\hline
\hline\\
\multicolumn{11}{c}{\textbf{\Large Best fitting parameters}} \\
 & & \multicolumn{4}{c}{\textbf{Hale - Best: Student's T}} &  & \multicolumn{4}{c}{\textbf{Anti-Hale - Best: von Mises}}\\ \cline{3-6} \cline{8-11}
\textbf{Fitted} & & & & & \textbf{Standard} & & & & \textbf{Standard} & \\
\textbf{Distribution} & & \textbf{n} & \textbf{Location $\mu$} & \textbf{Scale $s$} & \textbf{Deviation} & & \textbf{$\kappa$} & \textbf{Location $\mu$} & \textbf{Deviation} & \\
\hline
Student's T \textbf{(HH)} & & 3.45 & 7.5$^o$ & 14.46$^o$& 22.31$^o$ & & & & & \\
von Mises \textbf{(AH)} & & & & & & & 2.62 & 2.85$^o$ & 41.55$^o$ & \\
\end{tabular}
\label{tab:akaike_scores}
\end{table*}

\section{Optimal analytic fit to tilt angle populations} \label{sec:globalDist}

We now identify the analytic distributions that best capture the populations of HH and AH tilt.   We use two criteria to make model discrimination   The first one is Akaike's information criterion (AIC; Akaike 1983\nocite{akaike1983}).  The AIC is a powerful tool for discriminating between different non-nested models by making an estimate of the expected relative distance between the fitted model and the unknown true mechanism that generated the observed data.  The AIC for a model $M_j$ is defined as:
\begin{equation}\label{Eq_AIC}
  \mathrm{AIC_j} = - 2 \mathcal{L}(M_j) - 2 n_j,
\end{equation}
where $\mathcal{L}(M_j)$ is the log-likelihood of model $M_j$:
\begin{equation}\label{Eq_LL}
  \mathcal{L}(M) = \sum_{i=1}^n \log(\mathrm{P}(D_i|M)).
\end{equation}
and $n_j$ the number of parameters of model $j$.  The model with the smallest AIC is chosen as the best.  Minimizing AIC looks for the model with the largest log-likelihood.  However, log-likelihood alone is not sufficient to discriminate between models because it is biased as an estimation of the model selection target.  This bias was found by Akaike (1983\nocite{akaike1983}) to be approximately equal to each model's number of parameters ($n$), and thus the presence of the second term in Eq.~\ref{Eq_AIC}.  Together, log-likelihood and $n$ are used to strike a balance between bias and variance (or the trade-off between under-fitting and over-fitting).

The relative nature of the AIC is better represented by calculating the relative AIC differences:
\begin{equation}\label{Eq_AICDel}
  \mathrm{\Delta^{AIC}_j} = \mathrm{AIC_j} - \min(\mathrm{AIC}).
\end{equation}
This in turn can be used to estimate the likelihood of a model given the data:
\begin{equation}\label{Eq_AICL}
  \mathcal{L}(M_j|D) \propto \exp\left(-\frac{ \mathrm{\Delta^{AIC}_j}}{2}\right),
\end{equation}
and use it to calculate the Akaike weights:
\begin{equation}\label{Eq_AICW}
  \mathrm{Aw_j} = \frac{\exp\left(-\frac{ \mathrm{\Delta^{AIC}_j}}{2}\right)}{\sum_{k=1}^K \exp\left(-\frac{ \mathrm{\Delta^{AIC}_k}}{2}\right)},
\end{equation}
which are a measure of the probability that the model $M_j$ is the best model given the data.  For more information about AIC we recommend the excellent book by Burnham \& Anderson (2002\nocite{burnham-anderson2002}).

It is very important to highlight that the significance of AIC is strongly dependent on an appropriate choice of models.  Applying AIC to a set of very poor models will always select one estimated to be the best (even though that model may still be poor in an absolute sense).  To account for that, we also use the one sided K-S test under the null hypothesis that the HH and AH populations come from a given analytical distribution.   Table \ref{tab:akaike_scores} shows the result of our fits.  We test all distributions with infinite or circular support in the SciPy stats package \citep{2020SciPy-NMeth}.

The best fit, by far, to the HH tilt angle population is a Student's T distribution.  However, it is also the only one for which the null hypothesis of the K-S test cannot be rejected with confidence.  We find this a difficult result to interpret.   The Student's T distribution is overwhelmingly used to estimate the uncertainty in the estimation of the mean of a population governed by a normal distribution, given a small sample of observations, when the true variance of the distribution is unknown.  It is defined as:
\begin{equation}
    f(x|n,\mu,s) = \frac{\Gamma\left(\frac{n+1}{2} \right)}{s \sqrt{n\pi} \Gamma\left(\frac{n}{2} \right)} \left(1 + \frac{(x-\mu)^2}{ns^2} \right)^{-\frac{n+1}{2}},
\end{equation}
where $\Gamma$ is the gamma function, $n$ is a shape parameter typically associated with the number of degrees of freedom of a T distribution, $\mu$ is the location parameter, and $s$ the scale parameter.   We could not find an example where the T distribution is used to describe a physical population nor a physical generative model for this distribution.   We thought perhaps that this outcome stems from the fact that we were folding BMRs from all latitudes, each maybe with a normal distribution, but found that the T distribution is also the best fit, by far, for every 5$^o$ latitudinal bin shown in Figure \ref{fig:joyslaw}.  One possible explanation could be that the interaction of the rising flux-tube that gives rise to BMRs is buffeted by a limited amount of effective interactions with helical convective turbulence.  These would act as the small sample that gives rise to a mean imparted tilt described by a distribution with heavier tails than a normal distribution; even if the buffeting itself would be normally distributed.   The optimal fitted parameters (see Table \ref{tab:akaike_scores}) are 3.45 degrees of freedom ($n$), which we speculate could be related to the effective number of interactions with turbulence during BMR rise; a location ($\mu$) of 7.5$^o$, which quantifies the systematic tilt presented by HH BMRs; and a scale ($s$) of 14.46$^o$ which translates to a standard deviation of 22.31$^o$.

In the case of AH regions, the best fit (by far) is a von Mises distribution (also known as the circular normal distribution), defined as:
\begin{equation}
    f(x|\kappa,\mu) = \frac{e^{\kappa \cos(x-\mu)}}{2\pi \mathrm{I_0}(\kappa)},
\end{equation}
where $\kappa$ is a scale parameter analogous to $\sigma^2$ in a normal distribution, $\mu$ is the location parameter, and $\mathrm{I_0}(\kappa)$ is the modified Bessel function of order 0.  The von Mises distribution has circular support ($[-\pi,\pi]$) so its better suited for angular random variables.  The optimal fitted parameters (see Table \ref{tab:akaike_scores}) are a location ($\mu$) of 2.85$^o$, which demonstrates that AH BMRs also have a hemispheric systematic inclination (although not as marked as HH regions); and a scale factor ($\kappa$) of 2.62 which translates to a standard deviation of 41.55$^o$.   Although AIC indisputably shows a von Mises distribution as the best fit to our AH data, the K-S test can only confidently reject the Cauchy distribution as a possible analytical fit to AH tilts.   All other distributions are still consistent enough with the data given the K-S test.  Having more cycles worth of AH observations would help give certainty to our results.   However, repeating the exercise for the individual 5$^o$ bins shown in Figure \ref{fig:joyslaw} has the same result as with the HH distribution.  In every single case the AIC finds the von Mises distribution to be the best distribution by far.

\begin{figure}[ht!]
\centering
\includegraphics[scale=0.35]{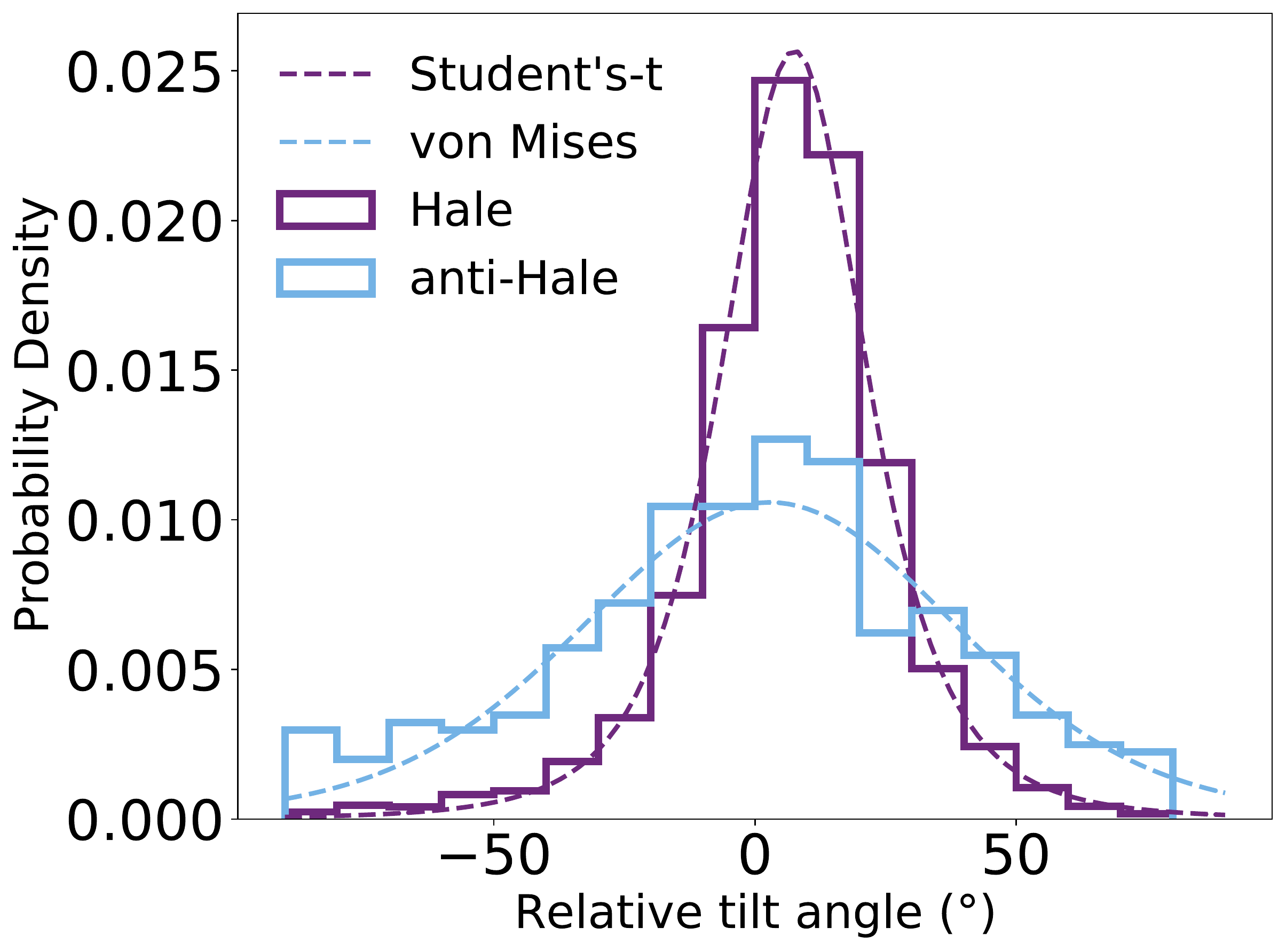}
\caption{Analytic fit to the relative tilt angle distribution for HH (dashed purple) and AH (dashed blue) populations.  The empirical distributions associated with each population are shown using the same colors, but a solid line . \label{fig:andresfinal}}
\end{figure}

Figure \ref{fig:andresfinal} shows the relative tilt angle distribution for HH and AH regions once all cycles and hemispheres have been combined, as well as the optimal fits to HH and AH BMR tilts.   The AH distribution is clearly more spread than the HH distribution, which is also evident in Figure \ref{fig:joyslaw} and Table \ref{tab:akaike_scores}.  It also shows how both distributions are shifted towards positive relative tilts, exhibiting the same hemispheric systematic tendency for HH and AH regions.

\begin{figure*}[ht!]
\includegraphics[width=\textwidth]{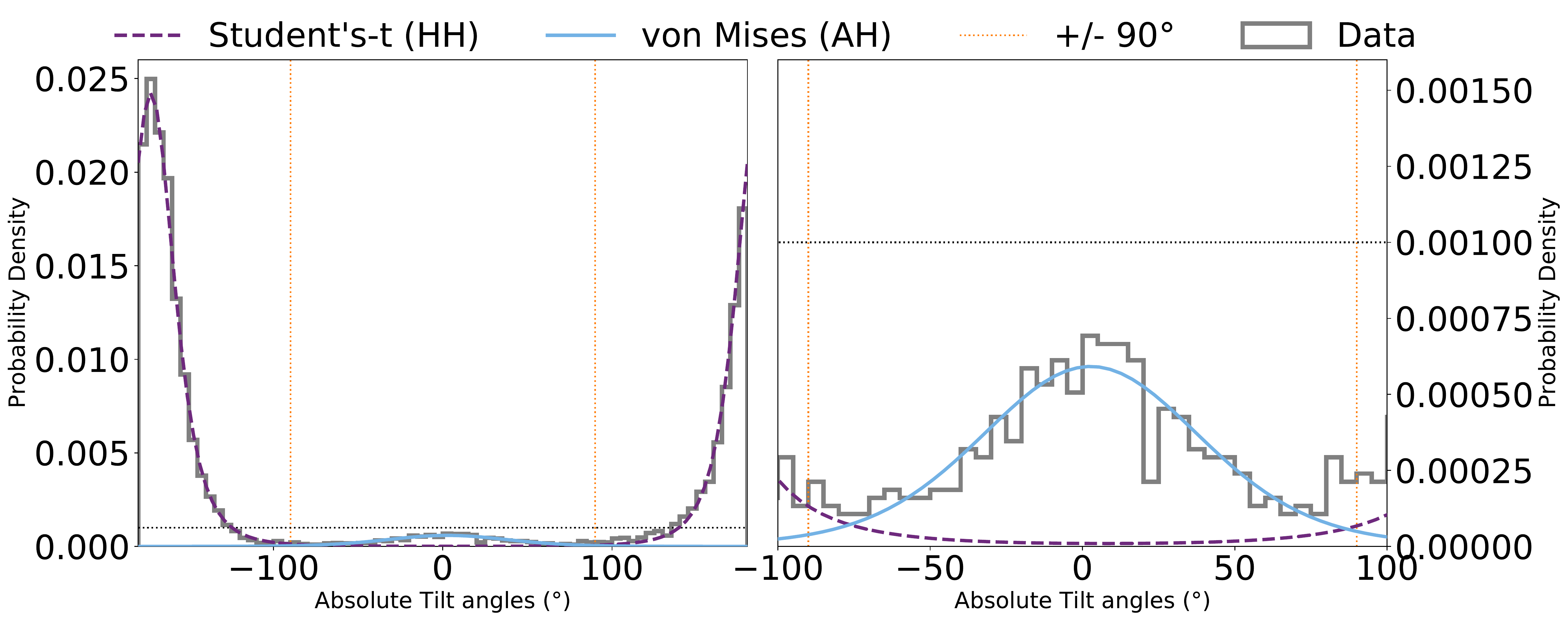}
\caption{\textit{Left:} Analytic fits of the von Mises (solid blue) plus Student's t (dashed purple) distributions to the absolute tilt of our entire database (solid gray).  Tilt angles have been shifted by 180$^0$ to place the focus of the plot on the AH regions. \textit{Right:} Zoom in on the AH peak.  The black dotted line is the same of both plots, as are the orange vertical dashed lines at $\pm$90$^o$.  \label{fig:tails}}
\end{figure*}

\section{The Anti-Hale population is not explained by Hale's tilt distribution tails} \label{sec:HaleDev}




Figure \ref{fig:tails} demonstrates that the tails of the HH distribution (dashed purple) cannot explain the relative frequency of AH regions, which needs to be explained by a second distribution (solid blue). The standard deviation associated with the HH and AH distributions indicate that in the vicinity of a tilt angle of $\pm$90$^o$ (dotted orange vertical lines), the majority of AH regions are likely to belong to the HH population.  However, the further from $\pm$90$^o$ the more likely that AH regions belong to the AH distribution. Around 14$^o$ past $\pm$90$^o$, AH regions are equally likely to belong to the HH and AH distributions, after which they are significantly more likely to belong to the AH distribution.

Given that the number of AH regions in our data is large enough (421), the results from both AIC and K-S rule out with a high degree of certainty that AH regions belong to the HH distribution.  We take the latter as evidence that AH regions constitute a distinct population and therefore their origin is different than the one of HH regions.  

\section{What is the origin of Anti-Hale regions?}
\label{sec:origin of AH regions}



The excess frequency of AH regions in the most equatorial bin (0-5$^o$), relative to the HH distribution (see Fig.~\ref{fig:joyslaw}), indicates that these equatorial AH regions are likely associated with the dominant toroidal flux system of the opposite hemisphere \citep[as reported by][]{mcclintock-etal2014}.

For all other bins (5-35$^o$) the latitudinal distribution of HH and AH regions are statistically indistinguishable.  We see this as evidence of a strong connection between them.  The systematic properties of AH regions suggest that they also originate from magnetic fields that are primarily toroidal (hence their systematic East-west orientation and systematically positive relative tilt.  We speculate that AH regions are connected to the prior emergence of HH regions with a very strong associated poloidal field contribution. This poloidal field could be sheared on the spot by differential rotation into a toroidal field (of the opposite polarity) that is strong enough to produce an AH region. This hypothesis may also explain why AH regions seem to be more frequent in stronger cycles (see Fig.~\ref{fig:porcentajes}), given that there is an increased likelihood of seeing large sunspot groups in stronger cycles \citet{munoz-etal2015b}. Our hypothesis could be validated if the emergence of AH regions is preceded consistently by HH regions with strong poloidal field contributions at a similar latitude and longitude.   If our hypothesis is true, AH regions would still be consistent with the existence of largely coherent toroidal flux systems inside the solar interior and may be playing an even more important role on the solar dynamo that is currently believed.

\section{Summary} \label{sec:conclusions}


We use a database consisting of 9,243 unique bipolar magnetic regions belonging to the last 4 solar Cycles (21, 22, 23, 24) measured by two ground-based instruments, KPVT/512 and KPVT/SPMG and two space-based instruments, SOHO/MDI and SDO/HMI.  Bipolar magnetic regions belonging to different instruments are calibrated by homogenizing BMR flux distributions and applying a flux threshold below which measurements of bipolar magnetic regions are likely to be strongly affected by instrument discrepancies.  We find that the only instrument that needs a calibration factor (0.8) is SOHO/MDI, which is in good agreement ($1/0.8 = 1.25$) with the 1.3 factor found by \cite{liu2012} to calibrate SDO/HMI to SOHO/MDI for strong magnetic fields, and adopt a flux threshold of $3\times 10^{21}$Mx.
This leaves 7,511 bipolar magnetic regions that we consider to be homogeneous enough to be studied together.  Out of this revised database, 421 (5.61\%) bipolar magnetic regions have anti-hale polarity orientation, which enables a robust statistical characterization that is not possible without the combination of observations from multiple solar cycles.

Our results show that we can reject, with a very high degree of certainty, that anti-Hale bipolar magnetic regions belong to the same population as Hale regions.  We also rule out, with a very high degree of certainty, that anti-Hale bipolar magnetic regions can be explained as Hale regions whose polarity orientation is an extreme realization of a unique physical mechanism.  Only anti-Hale regions that are oriented almost completely in the north-south direction (within 14$^o$ of a $\pm$90$^o$ fully north-south tilt) are likely to be flipped Hale regions.

Our results support the claim by \citet{stenflo-kosovichev2012} that anti-Hale and Hale regions do not have the same origin, and reject the suggestion by \citet{mcclintock-etal2014} that the anti-Hale regions can be accounted by the tails of the tilt angle distribution of the Hale bipolar magnetic regions.  However, we disagree with the interpretation of \citet{stenflo-kosovichev2012} that the existence of anti-Hale regions is in contradiction with the existence of a largely coherent toroidal flux system.

Our results show that Anti-Hale regions have a preferred East-West orientation and follow the same Joy hemispheric tendency as Hale regions: i.e.\ anti-Hale regions also have a tendency to have leading (trailing) polarities that are closer to the equator (pole).  This means that they originate from largely toroidal fields with a polarity orientation opposite to each hemisphere's dominant flux system.  We speculate that these toroidal fields are sheared locally ("on the spot") by differential rotation from the poloidal field associated with preceding Hale regions possessing a strong poloidal signature. This would explain why the relative frequency of anti-Hale regions seems to be larger for stronger cycles.  It also explains why they are observed coexisting with Hale regions at similar latitudes.  Finally, it also explains why anti-Hale regions have a latitudinal distribution that is statistically indistinguishable from that of Hale regions. This hypothesis can be validated (falsified) in future work by studying if there is a consistent presence of Hale regions with strong poloidal field contributions at similar latitude and longitude preceding anti-Hale emergence.  If our hypothesis is true, the generation of anti-Hale regions would be acting as a strong form of non-linear quenching mechanism for the solar cycle because they would counteract the strong positive contribution of the originating Hale regions.  It would also reconcile the existence of anti-Hale regions with the commonly held picture of a largely coherent toroidal flux system

The main results of this work are the following:
\begin{enumerate}
    \item The tilts of Bipolar Magnetic Regions from different cycles and hemispheres are consistent enough with each other that they can be considered to be from the same population and combined together.
    \item There seems to be monotonic relationship between cycle amplitude and the relative abundance of anti-Hale region emergence.
    \item The latitudinal distribution of Hale and anti-Hale regions can be considered to be the same, with the exception of the equatorial region (0-5$^o$) where anti-Hale regions are comparatively more frequent.
    \item Anti-Hale regions present the same systematic relative tilt angle inclination than Hale regions. i.e. anti-Hale regions' leading polarity tends to be closer to the equator than their trailing polarity.
    \item Anti-Hale region relative tilt also seems to increase with latitude, but this result may not be significant.
    \item Hale region relative tilt distribution is best fitted by a Student's T distribution centered on 7.5$^o$ and with a standard deviation of 22.31$^o$.
    \item Anti-Hale region relative tilt distribution is best fitted by a von Mises distribution centered on 2.85$^o$ and with a standard deviation of 41.55$^o$.
    \item Hale and anti-Hale regions cannot be considered to belong to the same population.
    \item Anti-Hale regions cannot be explained solely as Hale regions whose polarity has been flipped. i.e.\ as exceptional Hale regions.
\end{enumerate}

\section{Data Repository}
The database used for our analysis has been published by \cite{DVN/QEMSZ2_2021} in the Harvard Dataverse and can be accessed at \href{https://doi.org/10.7910/DVN/QEMSZ2}{https://doi.org/10.7910/DVN/QEMSZ2}.

\section{acknowledgments}
LEC acknowledges partial support from the Center of Excellence in Astrophysics and Associated Technologies (PFB06, CONICYT), and posthumously thanks Dr. Adelina Gutiérrez for her pioneering Solar Physics course (1971) at the Universidad de Chile.   This research was funded by NASA Living With a Star grant NNH15ZDA001N.  All our analysis was performed using the open-source NumPy \citep{2020NumPy} and SciPy \citep{2020SciPy-NMeth} packages. We thank Aimee Norton, an anonymous referee, and ApJ's statistics editor for very useful feedback and suggestions.



\bibliography{references_AMJ, References }
\bibliographystyle{aasjournal}



\end{document}